\documentclass[preprint,12pt]{elsarticle}%
\usepackage{amssymb}
\usepackage{amsfonts}
\usepackage{amsmath}
\usepackage{graphicx}
\usepackage{pgfplotstable}
\usepackage{siunitx}
\usepackage[T1]{fontenc}
\usepackage[lithuanian,english]{babel}
\usepackage[utf8x]{inputenc}%
\providecommand{\U}[1]{\protect\rule{.1in}{.1in}}
\begin{document}
\begin{frontmatter}
\title{Calculation of harmonic oscillator brackets in SU(3) basis}
\author[1]{Ramutis Kazys Kalinauskas}
\author[1]{Augustinas Step\v{s}ys\corref{cor1}}
\ead{augustinas.stepsys@ftmc.lt,augustinas.stepsys@gmail.com}
\cortext[cor1]{Corresponding author.}
\author[1]{Darius Germanas}
\author[2]{Saulius Mickevi\v{c}ius}
\address[1]{Center for Physical Sciences and Technology, Savanoriu ave. 231, LT-02300 Vilnius, Lithuania}
\address[2]{Vytautas Magnus University, K. Donelai\v{c}io str. 58, LT-44248, Kaunas, Lithuania}
\begin{abstract}

We present a new approach for the Talmi-Moshinsky transformation representation in the harmonic oscillator basis. We utilize the SU(3) scheme for calculation of harmonic oscillator brackets. Using this scheme we obtain the explicit relations for numeric evaluation and present a computational approach.    
\end{abstract}
\end{frontmatter}

\section{Introduction}

The harmonic oscillator basis (HO) is widely used for the approximation of the
wavefunction of the nuclear systems \cite{Navratil_2016,Johnson_2020}. Traditional ab initio approach for
constructing the model Hamiltonian depend on one-particle coordinates. As a
result, the model wavefunctions are not translationally invariant by default.
Hence, the wavefunction of such a system cannot be represented properly as
the center of mass (CM) coordinate of the nucleus has to be eliminated. An
additional procedure is required to solve this problem. However, this issue
can be avoided by the direct construction of the antisymmetric translationally
invariant wavefunction. In this approach, we use the intrinsic coordinates,
particularly, the normalized Jacobi coordinates,  by transitioning
from the one-particle to Jacobi coordinates you get the explicit removal of
the CM coordinate \cite{BARRETT2013131,PhysRevC.61.044001}.

While antisymmetrizing the wavefunction one has  to act upon the Jacobi
coordinates thus it is beneficial that the different sets of coordinates are
related to each other by the orthogonal transformations. The HO basis allows
to represent these transformations with a finite number of terms. The key
aspect of this approach is the Talmi-Moshinsky transformation and the
corresponding brackets (HOB) in the HO basis \cite{KAMUNTAVICIUS2001191}.

HOBs are frequently used in nuclear calculations. So, it is advantageous 
to have an efficient computational approach for the HOB calculation and there is a notable effort to produce new aproaches for this problem \cite{EFROS2021108005,EFROS2023108852,FATAH2025109409}.

In our previous papers, we devoted a significant effort to research HOBs.
Particularly, we presented the expressions of HO transformations for two and
three Jacobi coordinates, analyzed their symmetries, simplifications for
special cases of mass-ratio parameters \cite{MickeviciusGermanasKalinauskas+2012+421+428,STEPSYS201926}. Also, a set of computer codes were provided \cite{MICKEVICIUS20111377,GERMANAS2010420,STEPSYS2021108023,STEPSYS20143062,GERMANAS2017259}.

In this paper, we will present a different approach for HOB calculation.

\section{Definitions and motivation: HOB in SO(3) basis}

In the nuclear structure calculations, the single particle wavefunctions are
usually expanded in the HO eigenfunctions. Let us define the HO
function as
\begin{equation}
\left\langle \boldsymbol{r}|elm\right\rangle =\phi_{elm}\left(  \boldsymbol{r}%
\right)  =R_{el}(|r|)Y_{em}(\theta\phi),
\end{equation}
where $R_{el}(|r|)$ is the radial part and $Y_{em}(\theta\phi)$ is the
spherical harmonics. One of the possible function of HO is%
\begin{equation}
\phi_{elm}(\boldsymbol{r})=\left(  -1\right)  ^{n}\left(  \frac{2(n!)}%
{\Gamma(n+l+\frac{3}{2})}\right)  ^{1/2}\exp\left(  -\frac{r^{2}}{2}\right)
r^{l}L_{n}^{(l+\frac{1}{2})}\left(  r^{2}\right)  Y_{lm}(\frac{\boldsymbol{r}%
}{r}). \label{10}%
\end{equation}
In the eq.(\ref{10}) the corresponding dimensionless eigenvalue equals
$(e+\frac{3}{2})$; the principal quantum number $n=\frac{(e-l)}{2}=0,1,2,...$,
where $e=2n+l$ and $l$ is the angular momentum. $\Gamma$ denotes the gamma
function, $L_{n}^{(l+\frac{1}{2})}$is generalized Laguerre polynomial, and $Y$
is the spherical harmonics.
We define the two-particle wavefunction with bound momentum as
\begin{gather}
\{\phi_{e_{1}l_{1}}(\boldsymbol{r}_{1})\otimes\phi_{e_{2}l_{2}}(\boldsymbol{r}%
_{2})\}_{LM}=\left\langle \boldsymbol{r}_{1},\boldsymbol{r}_{2}|e_{1}%
l_{1},e_{2}l_{2}:LM\right\rangle \nonumber \\
=%
{\textstyle\sum\limits_{m_{1}m_{2}}}
\langle l_{1}m_{1},l_{2}m_{2}|LM\rangle\phi_{e_{1}l_{1}m_{1}}(\boldsymbol{r}%
_{1})\phi_{e_{2}l_{2}m_{2}}(\boldsymbol{r}_{2}),
\end{gather}
where $\langle l_{1}m_{1},l_{2}m_{2}|LM\rangle$ is the $SU(2)$ Clebsch-Gordan
coefficient. The transformation bracket for two HOs is defined as%
\begin{equation}
\left\vert e_{1}l_{1},e_{2}l_{2}:LM\right\rangle =%
{\displaystyle\sum\limits_{e_{1}^{\prime}l_{1}^{\prime},e_{2}^{\prime}%
l_{2}^{\prime}}}
\left\langle e_{1}l_{1},e_{2}l_{2}:L|e_{1}^{\prime}l_{1}^{\prime}%
,e_{2}^{\prime}l_{2}^{\prime}:L\right\rangle _{d}\left\vert e_{1}^{\prime
}l_{1}^{\prime},e_{2}^{\prime}l_{2}^{\prime}:LM\right\rangle 
\end{equation}
in \cite{KAMUNTAVICIUS2001191}. Here $d$ is a nonnegative real number, parameterizing the Talmi-Moshinsky transformation (see eq.(\ref{dmatrix})). The variables in the brackets are arranged in a fixed way and the orthogonal
transformation of the coordinates $\boldsymbol{r}_{1},\boldsymbol{r}_{2}$ is
written as the two-dimensional matrix%
\begin{equation}
\langle\boldsymbol{r}_{1}^{\prime},\boldsymbol{r}_{2}^{\prime}%
|\ \boldsymbol{r}_{1},\boldsymbol{r}_{2}\rangle_{d}=%
\left(
\begin{tabular}{ll}
$\sqrt{\frac{d}{1+d}}$ & $\phantom{-}\sqrt{\frac{1}{1+d}}$\\
$\sqrt{\frac{1}{1+d}}$ & $-\sqrt{\frac{d}{1+d}}$
\end{tabular}
\right). \label{dmatrix}%
\end{equation}
 This two-dimensional matrix is called the Talmi-Moshinsky transformation. In \cite{KAMUNTAVICIUS2001191}
the original expression for HOBs derived by B.Buck and A.C.Merchant \cite{BUCK1996387}
\begin{gather}
\left\langle EL,el:\Lambda\right.  \left\vert e_{1}l_{1},e_{2}l_{2}%
:\Lambda\right\rangle _{d}\nonumber\\
=i^{-\left(  l_{1}+l_{2}+L+l\right)  }\times2^{-\left(  l_{1}+l_{2}%
+L+l\right)  /4}\nonumber\\
\times\sqrt{\left(  n_{1}\right)  !\left(  n_{2}\right)  !\left(  N\right)
!\left(  n\right)  !\left[  2\left(  n_{1}+l_{1}\right)  +1\right]  !!\left[
2\left(  n_{2}+l_{2}\right)  +1\right]  !!\left[  2\left(  N+L\right)
+1\right]  !!\left[  2\left(  n+l\right)  +1\right]  !!}\nonumber\\
\times\sum_{abcdl_{a}l_{b}l_{c}l_{d}}\left(  -1\right)  ^{l_{a}+l_{b}+l_{c}%
}2^{\left(  l_{a}+l_{b}+l_{c}+l_{d}\right)  /2}d^{\left(  2a+l_{a}%
+2d+l_{d}\right)  /2}\left(  1+d\right)  ^{-\left(  2a+l_{a}+2b+l_{b}%
+2c+l_{c}+2d+l_{d}\right)  /2}\nonumber\\
\times\frac{\left[  \left(  2l_{a}+1\right)  \left(  2l_{b}+1\right)  \left(
2l_{c}+1\right)  \left(  2l_{d}+1\right)  \right]  }{a!b!c!d!\left[  2\left(
a+l_{a}\right)  +1\right]  !!\left[  2\left(  b+l_{b}\right)  +1\right]
!!\left[  2\left(  c+l_{c}\right)  +1\right]  !!\left[  2\left(
d+l_{d}\right)  +1\right]  !!}\nonumber\\
\times\left\{
\begin{array}
[c]{ccc}%
l_{a} & l_{b} & l_{1}\\
l_{c} & l_{d} & l_{2}\\
L & l & \Lambda
\end{array}
\right\}  \left\langle l_{a}0l_{c}0\right\vert \left.  L0\right\rangle
\left\langle l_{b}0l_{d}0\right\vert \left.  l0\right\rangle \left\langle
l_{a}0l_{b}0\right\vert \left.  l_{1}0\right\rangle \left\langle l_{c}%
0l_{d}0\right\vert \left.  l_{2}0\right\rangle \label{22}%
\end{gather}
was rewritten to make it more computationally feasible and symmetric. The
result took the shape of%
\begin{align}
&  \left\langle EL,el:\Lambda\right.  \left\vert e_{1}l_{1},e_{2}l_{2}%
:\Lambda\right\rangle _{d}\nonumber\\
&  =d^{\left(  e_{1}-e\right)  /2}\left(  1+d\right)  ^{-\left(  e_{1}%
+e_{2}\right)  /2}\sum_{e_{a}l_{a}e_{b}l_{b}e_{c}l_{c}e_{d}l_{d}}\left(
-d\right)  ^{e_{d}}\left\{
\begin{array}
[c]{ccc}%
l_{a} & l_{b} & l_{1}\\
l_{c} & l_{d} & l_{2}\\
L & l & \Lambda
\end{array}
\right\} \nonumber\\
&  \times G\left(  e_{1}l_{1};e_{a}l_{a},e_{b}l_{b}\right)  G\left(
e_{2}l_{2};e_{c}l_{c},e_{d}l_{d}\right)  G\left(  EL;e_{a}l_{a},e_{c}%
l_{c}\right)  G\left(  el;e_{b}l_{b},e_{d}l_{d}\right)  . \label{23}%
\end{align}
$G\left(  e_{1}l_{1};e_{a}l_{a},e_{b}l_{b}\right)  $ is expressed as trinomial
coefficients
\begin{align}
G\left(  e_{1}l_{1};e_{a}l_{a},e_{b}l_{b}\right)   &  =\sqrt{\left(
2l_{a}+1\right)  \left(  2l_{b}+1\right)  }\left\langle l_{a}0l_{b}%
0\right\vert \left.  l_{1}0\right\rangle \nonumber\\
&  \times\left[  \left(
\begin{array}
[c]{c}%
e_{1}-l_{1}\\
e_{a}-l_{a};e_{b}-l_{b}%
\end{array}
\right)  \left(
\begin{array}
[c]{c}%
e_{1}+l_{1}+1\\
e_{a}+l_{a}+1;e_{b}+l_{b}+1
\end{array}
\right)  \right]  ^{1/2}. \label{24}%
\end{align}
An additional component in the eq.(\ref{23}) is the 9j symbol, denoted
by the curly brackets. The 9j symbol in the expression has eight summation
indices, although only three are independent. The described expression
has proven to be useful in the ab initio nuclear structure calculations \cite{PhysRevC.108.054005,PhysRevC.109.014616,Htun2021,Le2020,PhysRevC.100.044313}. It is
well-balanced and gives fast and precise results. However, the calculation of the 9j coefficient is
still expensive, so an alternative expression that could produce nice results
would be useful.

\section{An alternative expression: HOB in $SU(3)$ basis}

In this section, we formulate the motivation for the alternative expression in
the $SU(3)$ basis. In quantum mechanics, the wavefunction is a scalar
product of two vectors in different bases in the Hilbert space. For two HOs we
can write the wavefunction as
\begin{equation}
\varphi_{e_{1}l_{1},e_{2}l_{2}:LM}(\boldsymbol{r}_{1},\boldsymbol{r}%
_{2})\equiv\{\phi_{e_{1}l_{1}}(\boldsymbol{r}_{1})\otimes\phi_{e_{2}l_{2}%
}(\boldsymbol{r}_{2})\}_{LM}\equiv\left\langle \boldsymbol{r}_{1}%
,\boldsymbol{r}_{2}|e_{1}l_{1},e_{2}l_{2}:LM\right\rangle .
\end{equation}
The intrinsic symmetry of a quantum three-dimensional HO is $U(3).$ If we have
two three-dimensional HOs then the group describing the system is $U(6).$ Lets mark the
symmetric irreducible representation (irrep) as $[E]$. $E$ is the total HO
energy for the system, ergo $E=e_{1}+e_{2}$. The basis vector as%
\begin{equation}
|e_{1}l_{1}e_{2}l_{2}:LM\rangle.
\end{equation}

This is the conventional basis for this problem we introduced in the
previous section. The group chain for this basis is

\begin{equation}
\begin{array}
[c]{ccccc}%
U(6) & \supset & U(3) & + & U(3)\\
&  & \cup &  & \cup\\
&  & SO(3) &  & SO(3)
\end{array}
.
\end{equation}

Here the quantum numbers for the irrep labelling are $e_{1},$ $e_{2}$ and
$l_{1},l_{2}.$ The angular momenta $l_{1}$ and $l_{2}$ are coupled to the
total angular momentum $L.$ Quantum numbers $e_{1},$ $e_{2}$ are the energy of
the HOs. However, we can try to use a different group chain for the
$U(6)$, namely%

\begin{equation}
\begin{array}
[c]{ccccc}%
U(6) & \supset & U(3) & \times & U(2)\\
&  & \cup &  & \cup\\
&  & SO(3) &  & SO(2)
\end{array}
.
\end{equation}
For this group chain the basis is%
\begin{equation}
|e_{1}e_{2}[E_{1}E_{2}]\alpha LM\rangle. \label{su_3_baze_0}%
\end{equation}
The quantum numbers $e_{1},$ $e_{2}$ are the energy of the HOs, $[E_{1}E_{2}]$
label the irreps of $U_{3}$, $L$ is the angular momentum and $M$ is the
projection of the angular momentum along the z-axis. $\alpha$ is the index of
recurrence of $L$ in the irrep $[E_{1}E_{2}]$ .

The basis in eq.(\ref{su_3_baze_0}) can also be written as%
\begin{equation}
|e_{1}e_{2}J\alpha LM\rangle\label{su_3_baze}%
\end{equation}
if we denote $\frac{E_{1}-E_{2}}{2}$ as $J$. The group chain $SU(3)\supset
SO(3)\supset SO(2)$ is a non-canonical subgroup chain.

A very important tool for coupling two $SU(3)$ representations are the
Clebsch-Gordan coefficients, or their isofactors. These coefficients emerge
when we do a reduction of a product of representations of $SU(3)$ into the
irreps of $SO(3).$ The correct combination of the quantum states from both
groups must be ensured. For the two irreducible representations $(\lambda
_{1},\mu_{1})$ and $(\lambda_{2},\mu_{2})$ of $SU(3)$ the decomposition is%

\begin{equation}
\{\lambda_{1},\mu_{1}\}\otimes\{\lambda_{2},\mu_{2}\}=\sum_{\lambda^{\prime
},\mu^{\prime}}C_{(\lambda^{\prime},\mu^{\prime})}^{(\lambda_{1},\mu
_{1}),(\lambda_{2},\mu_{2})}.
\end{equation}
The irreps of $SO(3)$ are marked as $\{\lambda^{\prime},\mu^{\prime
}\}$ $\ $ and the Clebsch-Gordan coefficients $C_{(\lambda^{\prime},\mu
^{\prime})}^{(\lambda_{1},\mu_{1}),(\lambda_{2},\mu_{2})}$ are the isofactors.
In agreement with Racah lemma \cite{Racah1942} the matrix elements
\begin{equation}
\langle(e_{1}l_{1},e_{2}l_{2})L|(e_{1},e_{2})[E_{1}E_{2}]\alpha L\rangle
\equiv\langle e_{1}l_{1},e_{2}l_{2}:L|e_{1}e_{2}J\alpha L\rangle
\end{equation}
that transform the vectors from one basis to the vectors of the second basis
are isofactors of Clebsch-Gordan coefficients of the group $SU(3).$ These
coefficients are used for the non-canonical embedding of the irreps. They can
also be denoted as $\left[
\begin{array}
[c]{ccc}%
e_{1} & e_{2} & j\alpha\\
l_{1} & l_{2} & L
\end{array}
\right]  $ or $C_{l_{1}l_{2}L}^{e_{1}e_{2},j\alpha}$ or $\langle l_{1}l_{2}|\alpha j\rangle^{e_{1}e_{2}L}$. We will use the last notation.

\section{The Talmi-Moshinsky transformation in $SU(3)$
representation}

In the first section, we presented the Talmi-Moshinsky transformation as the
two-dimensional matrix
\begin{equation}
\langle\boldsymbol{r}_{1}^{\prime},\boldsymbol{r}_{2}^{\prime}%
|\ \boldsymbol{r}_{1},\boldsymbol{r}_{2}\rangle_{d}=%
\left(
\begin{tabular}{ll}
$\sqrt{\frac{d}{1+d}}$ & $\phantom{-}\sqrt{\frac{1}{1+d}}$\\
$\sqrt{\frac{1}{1+d}}$ & $-\sqrt{\frac{d}{1+d}}$
\end{tabular}
\right). \label{dmatrix2}%
\end{equation}
This matrix has a determinant $-1$ and is a representation of the group
$O(2)$. Since $O(2)\supset SO(2)$, we rewrite the Talmi-Moshinsky (TM)
transformation as a representation of the $R$ $\times$ $SO(2)$, where $R$ is
reflection part of the transformation%

\begin{equation}
\left(
\begin{tabular}{ll}
$\sqrt{\frac{d}{1+d}}$ & $\phantom{-}\sqrt{\frac{1}{1+d}}$\\
$\sqrt{\frac{1}{1+d}}$ & $-\sqrt{\frac{d}{1+d}}$\\
\end{tabular}
\right)=
\left(
\begin{tabular}{ll}
$1$ & $\phantom{-}0$\\
$0$ & $-1$\\
\end{tabular}
\right)
 \times%
\left(
\begin{tabular}{ll}
$\phantom{-}\sqrt{\frac{d}{1+d}}$ & $\sqrt{\frac{1}{1+d}}$\\
$-\sqrt{\frac{1}{1+d}}$ & $\sqrt{\frac{d}{1+d}}$\\
\end{tabular}
\right) .
\end{equation}
Further, we will investigate the TM transformation without the reflection.
We write the TM as a general linear transformation matrix
\begin{equation}
\widehat{a}=
\left(
\begin{tabular}{ll}
$a_{1}$ & $a_{3}$\\
$a_{2}$ & $a_{4}$\\
\end{tabular}
\right) \label{matrix_aa}%
\end{equation}
and impose the properties of unitarity and unimodularity- $\widehat{a}%
\widehat{a}^{+}=1$ and $\det\widehat{a}=1.$ Immediately, we see that
$a_{2}=-a_{3}^{\ast}$ and $a_{4}=a_{1}^{\ast}$ producing%

\begin{equation}
\widehat{a}=%
\left(
\begin{tabular}{ll}
$\phantom{-}a_{1}$ & $a_{3}$\\
$-a_{3}^{\ast}$ & $a_{1}^{\ast}$\\
\end{tabular}
\right) , \label{matrix_a}%
\end{equation}
parametrized by the Cayley-Klein parameters \cite{Varshalovich1988}. The matrix (\ref{matrix_a}) is the representation of the $SU(2),$ which is the double covering group of $SO(2).$ We can specify the
matrix (\ref{matrix_a}) elements as

\begin{equation}%
\left(
\begin{tabular}{ll}
$\phantom{-}\cos(\theta)$ & $\sin(\theta)$\\
$-\sin(\theta)$ & $\cos(\theta)$\\
\end{tabular}
\right). \label{alphamatrix}%
\end{equation}

In  the matrix (\ref{alphamatrix}) TM transformation is parametrized with an angle $\theta,$
which is connected with $d$ in a straightforward way
\begin{gather}
\frac{1}{1+d}=\sin^{2}\theta,\\
\frac{1}{\sin^{2}\theta}=1+d,\\
d=\frac{1}{\sin^{2}\theta}-1=\frac{\sin^{2}\theta+\cos^{2}\theta-\sin
^{2}\theta}{\sin^{2}\theta}=\cot^{2}\theta.
\end{gather}

We need to find a representation of the transformation matrix
(\ref{alphamatrix}) in a $SU(3)$ basis. This orthogonal transformation must
transform the function of the coupled $SU(3)$ irreps $e_{1}$ and $e_{2}%
$, depending on the variables $\boldsymbol{r}_{1}$ and $\boldsymbol{r}_{2}%
$. The action of the transformation would result in a function depending on
the transformed variables $\boldsymbol{r}_{1}^{\prime}$ and $\boldsymbol{r}%
_{2}^{\prime}.$ We write the wavefunction of two coupled HOs in $SU(3)$ as
$\Psi(e_{1}e_{2}[E_{1}E_{2}]\gamma|\boldsymbol{r}_{1},\boldsymbol{r}_{2})$ or
in a bra-ket notation as $\langle\boldsymbol{r}_{1},\boldsymbol{r}_{2}%
|e_{1}e_{2}[E_{1}E_{2}]\gamma\rangle,$ where $\gamma=\alpha LM.$

If the ket vector is transformed by some operator $T$%
\begin{equation}
\langle\phi|T|\psi\rangle,
\end{equation}
then to keep the scalar product invariant, we need to act on a bra vector with
a some sort of operator $X$ in a such way that
\begin{equation}
X\langle\phi|T|\psi\rangle=\langle\phi|\psi\rangle.
\end{equation}
This procedure implies that the operator $X$ must satisfy the equality
\begin{equation}
X\langle\phi|=\langle\phi|T^{\dagger}.
\end{equation}
Thus, the operator $X$ is the Hermitian conjugate of the operator $T$ and can
be denoted as $T^{\dagger}$. We follow this outline and insert the
transformation $\hat{a}$ and its complex conjugate $\hat{a}^{\dagger}=\hat
{a}^{-1}$ in our bra-ket $\langle\boldsymbol{r}_{1},\boldsymbol{r}%
_{2}\left\vert \hat{a}^{-1}\hat{a}\right\vert e_{1}e_{2}[E_{1}E_{2}%
]\gamma\rangle$. We introduce the intermediate summation of $\frac
{(e_{1}^{\prime}-e_{2}^{\prime})}{2}$ and integrate out the operator $a^{-1}$
using the completeness relation for discrete and continuous spectra%

\begin{gather}
\langle\boldsymbol{r}_{1},\boldsymbol{r}_{2}\left\vert \hat{a}^{-1}\hat
{a}\right\vert e_{1}e_{2}[E_{1}E_{2}]\gamma\rangle\label{Dmatrix1}
=\int\int d\boldsymbol{r}_{1}d\boldsymbol{r}_{2}\langle\boldsymbol{r}%
_{1},\boldsymbol{r}_{2}|\hat{a}^{-1}|\boldsymbol{r}_{1}^{\prime}%
,\boldsymbol{r}_{2}^{\prime}\rangle \nonumber\\
\times {\textstyle\sum\limits_{(e_{1}^{\prime}-e_{2}^{\prime})/2}}
\langle\boldsymbol{r}_{1}^{\prime},\boldsymbol{r}_{2}^{\prime}|e_{1}^{\prime
}e_{2}^{\prime}[E_{1}E_{2}]\gamma\rangle\langle e_{1}^{\prime}e_{2}^{\prime
}[E_{1}E_{2}]\gamma|\hat{a}|e_{1}e_{2}[E_{1}E_{2}]\gamma\rangle\nonumber\\
=
{\textstyle\sum\limits_{(e_{1}^{\prime}-e_{2}^{\prime})/2}}
\langle\boldsymbol{r}_{1}^{\prime},\boldsymbol{r}_{2}^{\prime}|e_{1}^{\prime
}e_{2}^{\prime}[E_{1}E_{2}]\gamma\rangle\langle e_{1}^{\prime}e_{2}^{\prime
}[E_{1}E_{2}]\gamma|\hat{a}|e_{1}e_{2}[E_{1}E_{2}]\gamma\rangle\nonumber\\
=\Delta_{0}^{E_{2}}%
{\textstyle\sum\limits_{(e_{1}^{\prime}-e_{2}^{\prime})/2}}
\Psi(e_{1}^{\prime}e_{2}^{\prime}[E_{1}E_{2}]\gamma|\boldsymbol{r}_{1}%
^{\prime},\boldsymbol{r}_{2}^{\prime})d_{(e_{1}^{\prime}-e_{2}^{\prime
})/2,(e_{1}-e_{2})/2}^{(E_{1}-E_{2})/2}(\theta).
\end{gather}

The action of the operator $\hat{a}$ results in a Wigner d-matrix of the
irreducible representation of the group $SU(2)$, which is a subgroup of $SU(3)$. It is an orthogonal
transformation that transforms a function of the coupled $SU(3)$ irreps
$e_{1}$ and $e_{2}$ depending on the variables $\boldsymbol{r}_{1}$ and
$\boldsymbol{r}_{2}$ into a function depending on the variables
$\boldsymbol{r}_{1}^{\prime}$ and $\boldsymbol{r}_{2}^{\prime}$. The Wigner
d-matrix depends on the same angle $\theta$. We change the notation a little
bit, following the eq.(\ref{su_3_baze}), namely
\begin{equation}
J=\frac{(E_{1}-E_{2})}{2};M=\frac{(e_{1}-e_{2})}{2};M^{\prime}=\frac
{(e_{1}^{\prime}-e_{2}^{\prime})}{2}.
\end{equation}
This results in more common expression for eq.(\ref{Dmatrix1})%
\begin{equation}
\langle\boldsymbol{r}_{1},\boldsymbol{r}_{2}\left\vert \hat{a}^{-1}\hat
{a}\right\vert e_{1}e_{2}[E_{1}E_{2}]\gamma\rangle=\Delta_{0}^{E_{2}}%
{\textstyle\sum\limits_{m^{\prime}}}
\Psi(e_{1}^{\prime}e_{2}^{\prime}[E_{1}E_{2}]\gamma|\boldsymbol{r}_{1}%
^{\prime},\boldsymbol{r}_{2}^{\prime})d_{M,M^{\prime}}^{J}(\theta).
\end{equation}

The different expressions for the Wigner d-matrix are well known. For example,
we can use the expression from \cite{Vanagas}.%

\begin{gather}
d_{M,M^{\prime}}^{J}(\theta)=\sum_{k}\frac{\sqrt{(J+M)!(J-M)!(J+M^{\prime
})!(J-M^{\prime})!}}{k!(J-M-k)!(J+M^{\prime}-k)!(k+M-M^{\prime})!} \nonumber\\
\times a_{1}^{J+M^{\prime}-k}a_{2}^{k}a_{3}^{k+M-M^{\prime}}a_{4}^{J-M-k}.
\label{21.15}%
\end{gather}

This is a general expression, which is parametrized by a matrix
(\ref{matrix_aa}). From bounded conditions of the matrix (\ref{matrix_a}) we get the
connections
\begin{align}
\text{ }a_{1}  &  =\cos(\theta),\\
a_{4}  &  =-\cos(\theta),\\
a_{2}  &  =a_{3}=\sin(\theta).
\end{align}
Inserting them into the eq.(\ref{21.15}) results in an expression%

\begin{align}
d_{M,M^{\prime}}^{J}(\theta)  &  =\sum_{k}\frac{\sqrt{(J+M)!(J-M)!(J+M^{\prime
})!(J-M^{\prime})!}}{k!(J-M-k)!(J+M^{\prime}-k)!(k+M-M^{\prime})!}\nonumber\\
&  \times-\cos(\theta)^{J+M^{\prime}-k}\sin(\theta)^{k}\sin(\theta
)^{k+M-M^{\prime}}\cos(\theta)^{J-M-k}.
\end{align}
We rearrange this expression to make it computationally more feasible by using
$\Gamma$ function instead of factorials and pulling out terms of equation
independent from the summation%

\begin{gather}
d_{M,M^{\prime}}^{J}(\theta)=\sqrt{\Gamma(J+M+1)\Gamma(J-M+1)\Gamma
(J+M^{\prime}+1)\Gamma(J-M^{\prime}+1)}\label{Djm_vanagas}\nonumber\\
\times\sum_{k}\frac{-\cos(\theta)^{2J+M^{\prime}-M-2k}\sin(\theta
)^{2k+M-M^{\prime}}}{\Gamma(k+1)\Gamma(J-M-k+1)\Gamma(J+M^{\prime}%
-k+1)\Gamma(k+M-M^{\prime}+1)}.
\end{gather}
The momentum $J$ belongs in the
interval $|m|\leqslant J\leqslant J_{\max},$ where the $J_{\max}$ is equal to $\frac
{E}{2}.$ Each $J$ has its own values of $L$ recurrence index $\alpha$. The
recurrence index $\alpha$, also known as the degeneracy index, in a $SU(3)$ irrep can be
determined by the Racah $d_{(\lambda,\mu)}^{L}$ formula \cite{RevModPhys.21.494}. The index
$(\lambda,\mu)$ is yet another marking of the $SU(3)$ irrep, which is
connected to the notation presented before. Since we use quantum numbers $E\alpha J$, we
rewrite the Racah formula 
\begin{equation}
d_{(\lambda,\mu)}^{L}=\left[  \frac{\lambda+\mu-L+2}{2}\right]  _{\geq
0}-\left[  \frac{\lambda-L+1}{2}\right]  _{\geq0}-\left[  \frac{\mu-L+1}%
{2}\right]  _{\geq0},
\end{equation}
where $[\dots]_{\geq0}$ is zero for any negative number and gives the
positive integer part otherwise. This results in the equation%

\begin{gather}
d_{(\lambda,\mu)}^{L}=\alpha_{EJ}^{L}=\left[  \frac{E_{1}-L+2}{2}\right]
_{\geq0}-\left[  \frac{2J-L+1}{2}\right]  _{\geq0}-\left[  \frac{E_{2}-L+1}%
{2}\right]  _{\geq0}\nonumber\\
=\left[  \frac{E/2+J-L+2}{2}\right]  _{\geq0}-\left[  \frac{2J-L+1}{2}\right]
_{\geq0}-\left[  \frac{E/2-J-L+1}{2}\right]  _{\geq0}.
\end{gather}
As a result we see that for the given $EJL$ values the index $\alpha$ runs
in the interval $1\leqslant\alpha\leqslant\alpha_{EJ}^{L}=\alpha_{0}$.
This is a crucial information for the required $SU(3)$ Clebsch-Gordan
coefficient isofactor calculation. If $\alpha
_{EJ}^{L}$ equals to zero$,$ then the $SU(3)$ irrep $EJL$ does not exist. We
can calculate the dimension of the irrep using the expression
\begin{equation}
\dim(\lambda,\mu)=(1+\lambda)(1+\mu)\left(  1+\frac{\lambda+\mu}{2}\right) 
\end{equation}
from \cite{Draayer1968}. which we also rewrite in the $E\alpha J$ notation as
\begin{equation}
\dim (Ej)=(1+2J)(1+E/2-J)\left(  1+\frac{E/2+J}{2}\right)  .
\end{equation}
One can check that we indeed have%
\begin{equation}
d_{(\lambda,\mu)}=\sum_{L}(2L+1)d_{(\lambda,\mu)}^{L}.
\end{equation}

We calculate the reflection $R$ from the matrix 
\begin{equation}
a=\frac{1}{\Delta_{0}}%
\left( 
\begin{tabular}{ll}
$a_{1}$& $a_{3}$\\
$a_{2}$ & $a_{4}$\\
\end{tabular}
\right)
\end{equation}
 and the representation $d_{M,M^{\prime}}^{J}(\theta).$ When the matrix is
given a form $%
\left(
\begin{tabular}{ll}
$1$& $\phantom{-}0$\\
$0$ & $-1$\\
\end{tabular}
\right)
,$ we see that its representation is $d_{M,M^{\prime}}^{J}=(-1)^{J+M}%
\delta(M,M^{\prime}).$

\section{The calculation of the HOB}

In this section, we gather the presented knowledge into a new approach for the
calculation of the HOB. HOB is defined as $\langle e_{1}l_{1},e_{2}%
l_{2}:L|e_{1}^{\prime}l_{1}^{\prime},e_{2}^{\prime}l_{2}^{\prime}:L\rangle
_{d}$, or $\langle e_{1}l_{1},e_{2}l_{2}:L|T(d)|e_{1}^{\prime}l_{1}^{\prime
},e_{2}^{\prime}l_{2}^{\prime}:L\rangle.$ We transform the $SO(3)$ basis into
the $SU(3)$ basis by using the\ Clebsch-Gordan isofactors and summing though
the intermediate states%

\begin{gather}
\langle e_{1}l_{1},e_{2}l_{2}:L|T(d)|e_{1}^{\prime}l_{1}^{\prime}%
,e_{2}^{\prime}l_{2}^{\prime}:L\rangle\label{expression}
=%
{\textstyle\sum\limits_{E_{1}E_{2}\alpha E_{1}^{\prime}E_{2}^{\prime}%
\alpha^{\prime}}}
\langle e_{1}l_{1},e_{2}l_{2}:L|e_{1},e_{2}[E_{1}E_{2}]\alpha:L\rangle \nonumber \\
\times \langle e_{1},e_{2}[E_{1}E_{2}]\alpha:L|T(d)|e_{1}^{\prime},e_{2}^{\prime}%
[E_{1}^{\prime}E_{2}^{\prime}]\alpha^{\prime}:L\rangle
\langle e_{1}^{\prime
},e_{2}^{\prime}[E_{1}^{\prime}E_{2}^{\prime}]\alpha^{\prime}|e_{1}^{\prime
}l_{1}^{\prime},e_{2}^{\prime}l_{2}^{\prime}:L\rangle\nonumber\\
{\textstyle\sum\limits_{JM\alpha J^{\prime}M^{\prime}\alpha^{\prime}}}
\langle e_{1}l_{1},e_{2}l_{2}:L|e_{1},e_{2}J\alpha:L\rangle\langle e_{1}%
,e_{2}JM\alpha:L|T(d)|e_{1}^{\prime},e_{2}^{\prime}J^{\prime}M^{\prime}%
\alpha^{\prime}:L\rangle \nonumber \\
\times \langle e_{1}^{\prime},e_{2}^{\prime}J^{\prime}%
\alpha^{\prime}:L|e_{1}^{\prime}l_{1}^{\prime},e_{2}^{\prime}l_{2}^{\prime
}:L\rangle\delta_{J,J^{\prime}}\text{ }\delta_{\alpha,\alpha^{\prime}%
}.
\end{gather}
We make this expression more compact rewriting in more favorable notation.%
\begin{gather}
\langle e_{1}l_{1},e_{2}l_{2}:L|T(d)|e_{1}^{\prime}l_{1}^{\prime}%
,e_{2}^{\prime}l_{2}^{\prime}:L\rangle \nonumber \\%
= {\textstyle\sum\limits_{JM\alpha J^{\prime}M^{\prime}\alpha^{\prime}}}
\langle l_{1}l_{2}|J\alpha\rangle^{e_{1}e_{2}L}d_{M,M^{\prime}}^{\ J}%
(d)\langle J^{\prime}\alpha^{\prime}|l_{1}^{\prime}l_{2}^{\prime}%
\rangle^{e_{1}^{\prime}e_{2}^{\prime}L}\delta_{\alpha,\alpha^{\prime}}%
\delta_{J,J^{\prime}}.
\end{gather}
By taking into account the $\delta$ functions we get the expression
\begin{equation}
\langle e_{1}l_{1}e_{2}l_{2}L|T(d)|e_{1}^{\prime}l_{1}^{\prime}e_{2}^{\prime
}l_{2}^{\prime}L\rangle=%
{\textstyle\sum\limits_{JMaM^{\prime}}}
\langle l_{1}l_{2}|J\alpha\rangle^{e_{1}e_{2}L}d_{M,M^{\prime}}^{\ J}%
(d)\langle J\alpha|l_{1}^{\prime}l_{2}^{\prime}\rangle^{e_{1}^{\prime}%
e_{2}^{\prime}L}.
\end{equation}
We further simplify this expression by using the relations 
\begin{gather}
e_{1}-e_{2}=2M;\\e_{1}^{\prime}-e_{2}^{\prime}=2M^{\prime};\\e_{1}+e_{2}%
=e_{1}^{\prime}+e_{2}^{\prime}=E.
\end{gather}
This results in the equation
\begin{gather}
\langle e_{1}l_{1}e_{2}l_{2}L|T(d)|e_{1}^{\prime}l_{1}^{\prime}e_{2}^{\prime
}l_{2}^{\prime}L\rangle= \nonumber \\
{\textstyle\sum\limits_{J=\max(|M|,|M^{\prime}|)}^{E/2}}
d_{M,M^{\prime}}^{\ J}(d)%
{\textstyle\sum\limits_{\alpha=1}^{\alpha_{0}}}
\langle l_{1}l_{2}|J\alpha\rangle^{e_{1}e_{2}L}\langle l_{1}^{\prime}%
l_{2}^{\prime}|J\alpha\rangle^{e_{1}^{\prime}e_{2}^{\prime}L}.
\label{final_expression}%
\end{gather}
We are left with two summations of permitted $J$ and $\alpha$ values with bounds
$|M|\ \leq J\leq E/2$ and $|M^{\prime}|\ \leq J\leq E/2$. The multiplicity label
$\alpha$ enumerates the reccuring $SO(3)$ irrep $L$ in the $SU(3)$ irrep $J$. The label $\alpha$ is the same as the label $K$ introduced by J. P. Elliott in \cite{Elliott1958}. 
The CG coefficients in the SO(3) bases are constructed subsequently from the
overlaps between the SU(2) and SO(3) bases.
\section{Computation of $SU(3)$ HOB}

The computation of the eq.(\ref{final_expression}) is straightforward.
For testing we used the matrix multiplication version, matrix element code, and compared the results to a HOB calculation code from \cite{KAMUNTAVICIUS2001191}. We employed Fortran90 language. For the $SU(3)\supset
SO(3)$ CG coefficient calculation we use the computer code created by C. Bahri, D.J. Rowe, and J.P. Draayer \cite{BAHRI2004121}.
The eq.(\ref{Djm_vanagas}) is computed using the standard Fortran Gamma
function. The calculation effort of a Wigner D matrix element is negligible
and all the heavy lifting is done by the CG coefficients. We used a GNU Fortran
compiler and the profiling was done using gprof. We checked the calculation
time for HO E=[0,20] with all the L's=[0,E] and found the CG coefficient
computation takes 82.46 \% of the time in function for matrix element
computation. The d-matrix element calculation takes only 0.51 \%
of all the calculation. The rest are auxiliary functions. We calculated the HOB
matrix for different E and L which results in dense matrices. The matrix for a
given E would have a block-diagonal structure depending on L. This structure
allows to build matrix for a given HO E by calculating only blocks. We sum
the time and the error for all the values of L for a given  HO E.

\begin{table}[h]
\centering
\begin{tabular}
[c]{|c|c|c|c|}\hline
\textbf{E} & \textbf{$\delta{(MatMult)}$} & \textbf{$\delta{(HOB)}$} &
\textbf{$\delta{(SU3HOB)}$}\\\hline
0 & 0.100E-04 & 0.100E-05 & 0.800E-05\\\hline
1 & 0.500E-05 & 0.100E-05 & 0.160E-04\\\hline
2 & 0.120E-04 & 0.400E-05 & 0.460E-04\\\hline
3 & 0.280E-04 & 0.150E-04 & 0.172E-03\\\hline
4 & 0.760E-04 & 0.580E-04 & 0.643E-03\\\hline
5 & 0.155E-03 & 0.151E-03 & 0.236E-02\\\hline
6 & 0.421E-03 & 0.491E-03 & 0.737E-02\\\hline
7 & 0.811E-03 & 0.117E-02 & 0.216E-01\\\hline
8 & 0.151E-02 & 0.264E-02 & 0.509E-01\\\hline
9 & 0.272E-02 & 0.533E-02 & 0.115E+00\\\hline
10 & 0.490E-02 & 0.113E-01 & 0.265E+00\\\hline
11 & 0.851E-02 & 0.247E-01 & 0.569E+00\\\hline
12 & 0.152E-01 & 0.457E-01 & 0.118E+01\\\hline
13 & 0.239E-01 & 0.839E-01 & 0.237E+01\\\hline
14 & 0.393E-01 & 0.151E+00 & 0.460E+01\\\hline
15 & 0.651E-01 & 0.291E+00 & 0.874E+01\\\hline
16 & 0.102E+00 & 0.469E+00 & 0.163E+02\\\hline
17 & 0.153E+00 & 0.799E+00 & 0.288E+02\\\hline
18 & 0.226E+00 & 0.131E+01 & 0.494E+02\\\hline
19 & 0.325E+00 & 0.209E+01 & 0.824E+02\\\hline
20 & 0.491E+00 & 0.346E+01 & 0.137E+03\\\hline
21 & 0.701E+00 & 0.548E+01 & 0.222E+03\\\hline
22 & 0.101E+01 & 0.851E+01 & 0.353E+03\\\hline
23 & 0.141E+01 & 0.132E+02 & 0.555E+03\\\hline
\end{tabular}
\caption{Calculation time in seconds of HOB block matrices for a given HO
Energy quanta. E is the HO Energy, Matmult- Matrix multiplication version of
the code, HOB code \cite{KAMUNTAVICIUS2001191}, SU3HOB- matrix element code.}%
\label{tab:time}%
\end{table}

In the table (\ref{tab:time}) we see that the calculation of the MatMult method is
faster by a factor of ten, than the HOB code from \cite{KAMUNTAVICIUS2001191}. The
SU3HOB matrix element code allows the computation of the HOB matrix by a factor of ten slower.

For the error calculation, we use the fact that the calculated matrices
must be orthonormal, that is  $HOBxHOB^{T}=I.$ Thus we can calculate the deviation from
the unitary matrix. We norm the result using the dimension of the matrix to
see the deterioration of the matrix as for larger matrices
we would have more error. We also sum for a given HO E. The normed
error values for MatMult and SU3HOB are the same. The results are presented in
table (\ref{tab:reduced_results}).

\begin{table}[h]
\centering
\begin{tabular}
[c]{|c|c|c|}\hline
\textbf{E} & \textbf{$\delta{(SU3HOB)}$} & \textbf{$\delta{(HOB)}$}\\\hline
0 & 0.143E-12 & 0.000E+00\\\hline
1 & 0.141E-12 & 0.444E-15\\\hline
2 & 0.141E-12 & 0.460E-15\\\hline
3 & 0.142E-12 & 0.791E-15\\\hline
4 & 0.143E-12 & 0.178E-14\\\hline
5 & 0.143E-12 & 0.372E-14\\\hline
6 & 0.145E-12 & 0.668E-14\\\hline
7 & 0.146E-12 & 0.166E-13\\\hline
8 & 0.162E-12 & 0.649E-13\\\hline
9 & 0.185E-12 & 0.153E-12\\\hline
10 & 0.424E-12 & 0.746E-12\\\hline
11 & 0.878E-12 & 0.231E-11\\\hline
12 & 0.201E-11 & 0.632E-11\\\hline
13 & 0.537E-11 & 0.327E-10\\\hline
14 & 0.269E-10 & 0.106E-09\\\hline
15 & 0.710E-10 & 0.505E-09\\\hline
16 & 0.262E-09 & 0.134E-08\\\hline
17 & 0.713E-09 & 0.503E-08\\\hline
18 & 0.189E-08 & 0.190E-07\\\hline
19 & 0.790E-08 & 0.683E-07\\\hline
20 & 0.328E-07 & 0.273E-06\\\hline
21 & 0.155E-06 & 0.924E-06\\\hline
22 & 0.687E-06 & 0.331E-05\\\hline
23 & 0.279E-05 & 0.133E-04\\\hline
\end{tabular}
\caption{Error of the HOB matrices calculated by HOB and SU3HOB for a given HO
Energy quanta. E is the HO Energy, HOB code from \cite{KAMUNTAVICIUS2001191}, SU3HOB- matrix element
code.}%
\label{tab:reduced_results}%
\end{table}

We see that for small HO E up until E=12, the HOB code is more accurate,
however for larger E, the SU3HOB takes the lead. More detailed analysis of
data shows that for larger matrices where L values are up until approximately
$\frac{E}{2},$ the HOB code deteriotes faster. However, for larger Ls, the
SU3HOB code gives worse results (see table (\ref{tab:reduced_dataset})). This can be
attributed to the SU3 Clebsch-Gordan code, which has trouble coupling for the
larger Ls. We checked the errors for the Clebsch-Gordan matrix and the d-matrix used for matrix multiplication code following the same condition of orthonormality. The results are presented in table (\ref{tab:data_values}). We see a clear deterioration of the CG matrix, while the d matrix is pretty stable.
\begin{table}[h]
\centering
\begin{tabular}
[c]{|c|c|c|c|c|}\hline
\textbf{E} & \textbf{L} & \textbf{Dimension} & \textbf{$\delta{(HOB)}$} &
\textbf{$\delta{(SU3HOB)}$}\\\hline
23 & 1 & 156 & 0.176E-04 & 0.118E-08\\\hline
23 & 2 & 132 & 0.124E-04 & 0.909E-09\\\hline
23 & 3 & 264 & 0.118E-03 & 0.337E-08\\\hline
23 & 4 & 220 & 0.110E-03 & 0.304E-08\\\hline
23 & 5 & 330 & 0.540E-03 & 0.442E-07\\\hline
23 & 6 & 270 & 0.480E-03 & 0.398E-07\\\hline
23 & 7 & 360 & 0.250E-02 & 0.155E-06\\\hline
23 & 8 & 288 & 0.159E-02 & 0.138E-06\\\hline
23 & 9 & 360 & 0.125E-01 & 0.533E-06\\\hline
23 & 10 & 280 & 0.472E-02 & 0.475E-06\\\hline
23 & 11 & 336 & 0.233E-01 & 0.118E-05\\\hline
23 & 12 & 252 & 0.414E-02 & 0.105E-05\\\hline
23 & 13 & 294 & 0.101E-01 & 0.234E-05\\\hline
23 & 14 & 210 & 0.881E-03 & 0.204E-05\\\hline
23 & 15 & 240 & 0.190E-02 & 0.155E-04\\\hline
23 & 16 & 160 & 0.955E-04 & 0.136E-04\\\hline
23 & 17 & 180 & 0.327E-04 & 0.878E-04\\\hline
23 & 18 & 108 & 0.214E-05 & 0.744E-04\\\hline
23 & 19 & 120 & 0.428E-06 & 0.431E-03\\\hline
23 & 20 & 60 & 0.166E-07 & 0.372E-03\\\hline
23 & 21 & 66 & 0.101E-08 & 0.297E-02\\\hline
23 & 22 & 22 & 0.450E-10 & 0.256E-02\\\hline
23 & 23 & 24 & 0.842E-11 & 0.668E-02\\\hline
\end{tabular}
\caption{Error of the HOB matrices calculated by 3HOB and SU3HOB for a given
HO Energy quanta. E is the HO Energy, 3HOB code from \cite{KAMUNTAVICIUS2001191}, SU3HOB- matrix
element code of derived expression.}%
\label{tab:reduced_dataset}%
\end{table}

\begin{table}[h]
\centering
\begin{tabular}
[c]{|c|c|c|}\hline
\textbf{E} & \textbf{$\delta{(CGMat)}$} & \textbf{$\delta{(Djmm^{\prime})}$%
}\\\hline
1 & 0.141E-12 & 0.444E-15\\\hline
2 & 0.493E-12 & 0.266E-14\\\hline
3 & 0.850E-12 & 0.344E-14\\\hline
4 & 0.186E-11 & 0.133E-13\\\hline
5 & 0.280E-11 & 0.231E-13\\\hline
6 & 0.496E-11 & 0.575E-13\\\hline
7 & 0.701E-11 & 0.754E-13\\\hline
8 & 0.114E-10 & 0.149E-12\\\hline
9 & 0.165E-10 & 0.200E-12\\\hline
10 & 0.521E-10 & 0.372E-12\\\hline
11 & 0.141E-09 & 0.483E-12\\\hline
12 & 0.411E-09 & 0.901E-12\\\hline
13 & 0.153E-08 & 0.126E-11\\\hline
14 & 0.935E-08 & 0.242E-11\\\hline
15 & 0.293E-07 & 0.373E-11\\\hline
16 & 0.133E-06 & 0.652E-11\\\hline
17 & 0.422E-06 & 0.952E-11\\\hline
18 & 0.136E-05 & 0.152E-10\\\hline
19 & 0.704E-05 & 0.219E-10\\\hline
20 & 0.371E-04 & 0.342E-10\\\hline
21 & 0.208E-03 & 0.495E-10\\\hline
22 & 0.111E-02 & 0.746E-10\\\hline
23 & 0.523E-02 & 0.111E-09\\\hline
\end{tabular}
\caption{Error of the Clebsch-Gordan matrix compared to Wigner d-matrix for a
given HO Energy quanta.}%
\label{tab:data_values}%
\end{table}

\section{Conclusions}

We have introduced a new approach for the harmonic
oscillator brackets using the SU(3) scheme. The SU(3) scheme allows a 
simple and elegant expression for the calculation of the Talmi-Moshinsky
transformation representation in the HO basis. We compared our calculation
time and errors to widely used code \cite{KAMUNTAVICIUS2001191}. While our matrix
element code lags behind the HOB code, we get more accuracy. When the whole
HOB matrix is needed for given HO E and L, the matrix multiplication approach
is superior from both accuracy and time consumption points of view. All the
heavy lifting in the calculation of the SU3HOB is done by the Clebsch-Gordan
coefficient calculation code \cite{BAHRI2004121}. Optimization of the said coefficient calculation
would allow us to get faster and maybe more accurate results.

Another important aspect is that in this scheme we can calculate the Wigner d-matrix with different kinds of d parameters for various Jacobi coordinate transformations and calculate the transformation
between the SO(3) and SU(3) basis only once. This approach allows an exciting
opportunity to formulate the ab initio no-core shell model in the unitary
scheme using Jacobi coordinates. 

\section*{Acknowledgements}
This project has received funding from the Research Council of Lithuania (LMTLT), agreement No S-PD-22-9\\
The authors gratefully acknowledge the computing time granted on the
supercomputer JURECA \cite{JURECA} at Forschungszentrum Juelich.

\bibliographystyle{elsarticle-num}
\bibliography{bibliography}

\end{document}